\documentstyle[twoside,fleqn,espcrc2_pre,espcrc2,epsf]{article}
 
\newcommand{\vsbt}{\vspace*{-0pt}}
\newcommand{\vsat}{\vspace*{-0pt}}
\newcommand{\vsbtc}{\vspace*{-0pt}}

\newcommand{\x}{\mbox{$\underline{x}$}}

\newcommand{\rr}{\underline{r}}
\newcommand{\MILCbeta}{\beta}
\newcommand{\rrmod}{\mbox{$\mid \underline{r} \mid$}}

\newcommand{\p}{\mbox{$\underline{p}$}}

\begin{document}
 
\title{Assorted  weak matrix elements involving the bottom quark\thanks{
Presented by C.\ McNeile.}
}

\author{C.\  Bernard,\address{Department of Physics, 
Washington University, St.\ Louis, MO 63130, USA}
        T.\ Blum,\address{Department of Physics,
Brookhaven National Lab, Upton, NY 11973, USA}
	T.\ DeGrand,\address{Physics Department,
University of Colorado, Boulder, CO 80309, USA}
	C.\ DeTar,\address{Physics Department,
University of Utah, Salt Lake City, UT 84112, USA}
	Steven Gottlieb,\address{Department of Physics,
Indiana University, Bloomington, IN 47405, USA}
	U.\ M.\ Heller,\address{SCRI, The Florida State University,
Tallahassee, FL 32306-4052, USA}
	J.\ Hetrick,${}^b$
	C.\ McNeile,${}^d$
	K.\ Rummukainen,${}^e$
	R.\ Sugar,\address{Department of Physics,
University of California, Santa Barbara, CA 93106, USA}
	D.\ Toussaint,\address{Department of Physics,
University of Arizona, Tucson, AZ 85721, USA}
	and
	M.\ Wingate${}^c$}

\begin{abstract}
As part of a larger project to estimate the 
$f_{B}$ decay constant, we are recalculating
$f_{B}^{static}$  using a variational smearing method
in an effort to improve accuracy.
Preliminary results for the static  $B_{B}$ parameter and  HQET 
two point functions are also presented.
\end{abstract}
 
\maketitle

\section{INTRODUCTION}
The extraction of CKM matrix elements from experimental data requires the 
calculation of QCD matrix elements of operators involving the 
bottom quark~\cite{FLYNN96a}.
For a number of years the MILC collaboration has been doing a
systematic study of the $f_{B}$
decay constant~\cite{MILC}. The difficulty of simulating the bottom quark, 
with its mass  larger than current feasible inverse lattice spacings, was overcome by
interpolating between the results of simulations using Wilson quarks with masses
around the charm mass and those from static simulations (infinite mass). 

Unfortunately, for some of the simulations  the $f_B^{static}$ results
were not useable.  These results came as a by-product of the
hopping parameter expansion of the heavy quark, and for technical
reasons had significant contamination by higher momentum intermediate
states when the physical volume was large~\cite{MILC}.  In addition, it is well
known that the poor signal to noise ratio of
the static simulations makes it important to use an efficient smearing method.
To provide $f_B^{static}$ results on all lattices,
and to reduce the errors even in the cases where the results were
previously available,  we have started a set of static-light simulations
using a variational smearing method.
\section{STATIC $f_{B}$ }
The computational cost of our current implementation of the FFT on  parallel machines,  prohibited the use of
a more sophisticated smearing technique such as MOST~\cite{Draper94a0}, so we  use
the following basis of smearing functions:
\begin{eqnarray}
 s(\rr)_{1} & = & e^{ -A * \rrmod }  \nonumber \\
 s(\rr)_{2} & = & e^{ -A * \rrmod } * ( 1 - B*\rrmod)  \nonumber \\
 s(\rr)_{3} & = & e^{ -A * \rrmod } * ( 1 - C*\rrmod - D * \rrmod^{2} )  \nonumber
\end{eqnarray} 
The parameters $A$, $B$, $C$ and $D$
were obtained from uncorrelated
fits to the Kentucky group's 
measured wave functions of static-light mesons 
(at $\MILCbeta=6.0$)~\cite{Draper94a0}.
The $A$, $B$, $C$ and $D$ parameters
were scaled to the appropriate value for a given $\MILCbeta$, using
the estimates of the lattice spacing.
Because of uncertainties in the lattice spacing, and to have the 
flexibility to choose different smearing functions for each kappa
value,
two additional sets of the parameters  were chosen.
Thus a variational smearing matrix of order ten (including the local
operator) is
used in the simulations.
The static quark is smeared {\em relative} to the light quark using
standard FFT methods.
%
%
\begin{table}[t]
\caption{Simulation parameters}
\label{tb:simres}
\vsbt
\centering
\begin{tabular}{ccccc} \hline
$\MILCbeta$ & volume &  \# configs & $a m_{sea}$  & \# $\kappa$  \\ \hline
  $5.6$    & $16^{3} \times 32$   & 100  & $0.01$   & 3  \\
  $5.5$    & $24^{3} \times 64$   & 100  & $0.1$    & 3 \\
 $5.445$  & $16^{3} \times 48$    & 100  & $0.025$   &  3 \\ \hline
\end{tabular}
   \medskip
   \vsbtc
   \vsat
\vspace{-0.3in}
\end{table}
The completed static-light production runs are shown
in Table~\ref{tb:simres}. The $\MILCbeta = 5.6$ configurations were
generated
by the HEMGCC collaboration~\cite{Bitar92a}.
Our analysis of the data is very
preliminary--in particular we have not yet fully optimized our smearing functions.
All the results presented here use a single exponential source.
We will focus on comparing the new
static results with the numbers from other simulations. Ultimately,
all the static data will be combined with that from propagating
quark simulations~\cite{MILC}.

We do a simultaneous fit to the smeared-local and 
smeared-smeared 
correlators. To compare our raw lattice numbers with
other simulations, we quote our numbers in terms of $Z_{L}$,
defined by
\begin{eqnarray}
C_{LS}(t)  & =  & Z_{L} Z_{S} e^{ -E_{sim} t} \\
C_{SS}(t)  & =  & Z_{S} Z_{S} e^{ -E_{sim} t} 
\end{eqnarray}
$Z_{L}$ is related to the decay constant, no perturbative factors are
included
in its definition and we assume the light quark propagator has been multiplied by $2 \kappa$.
$E_{sim}$ is the energy of the static-light meson, it is equal to the sum of the
difference between the mass of the $B$ meson and the bottom quark
mass, and an unphysical $\frac{1}{a}$ renormalization factor.
Correlations were included for the fits in time, but no kappa correlations
were included in the chiral extrapolations.

Some preliminary,
static fit results for the $\MILCbeta = 5.6$ simulation are
contained in Table \ref{tb:fiftysix}.
\begin{table}[t]
\caption{$Z_{L}$ fit results for $\MILCbeta = 5.6$, $a m_{sea} = 0.01$, with a
source $\exp{(-0.4 * \rrmod)}$, 
and fit region 7 to 11}
\label{tb:fiftysix}
  \centering
\begin{tabular}{cccc} \hline
 $\kappa$  & $a E_{sim}$ & $ a^{3/2} Z_{L} $  & $ \chi^{2}/dof $ \\ \hline
0.156   & $0.708_{-3}^{+6}$  & $0.312_{-4}^{+7}$ & $4.7/7$ \\
0.158   & $0.684_{-4}^{+7}$  & $0.287_{-5}^{+8}$ & $4.9/7$ \\
0.159   & $0.637_{-4}^{+7}$  & $0.274_{-5}^{+7}$ & $5.6/7$ \\ \hline
0.16103 & $0.649_{-5}^{+8}$  & $0.249_{-6}^{+7}$ & $0.01/1$     \\ \hline
\end{tabular}
  \medskip
  \vsbtc
  \vsat
\vspace{-0.3in}
\end{table}
Ali Khan {\it et al.}~\cite{AliKhan96a}
have also calculated $Z_{L}$ on gauge configurations from the same 
$\MILCbeta = 5.6$ simulation, as those used here, but not on
exactly the same sample of configurations. At $\kappa = 0.1585$,  they quote
$aE_{sim}=0.528(5)$ and $a^{3/2} Z_{L} = 0.24(3)$. Although we do
not have this kappa value in our
simulation we can use the information obtained in the chiral
fit model to estimate $a^{3/2} Z_{L} = 0.280^{+7}_{-5}$ and $aE_{sim} = 0.536^{+7}_{-4}$,
(where we have added $\log u_{0}$, with $u_{0} = 0.867$,
to our value of $a E_{sim}$ because Ali Khan {\it et al.}\  rescaled all their gauge fields
by the tadpole improvement factor of $u_{0}$).
The results from the older MILC static calculation also agree with those
from the new simulations.

We have done some fits to the $\beta=5.445$ static
data. The masses obtained were consistent with the older MILC
calculation. This is a large-volume  case where the older method
does not produce useable static $Z_L$ factors, so no check is available
there.  However, the new static
correlators seem reasonably consistent with the propagating quark
results.

\section{STATIC $B_{B}$ PARAMETER}
The $B_{B}$ parameter is required in the extraction of the 
$V_{td}$ CKM matrix element from the experimental
data on $\overline{B}-B$ mixing.
For the $\MILCbeta=5.5$ and $5.6$ simulations we
calculated the static $B_{B}$ parameter (so far we have only analyzed the
$\MILCbeta = 5.6$ data).
Ours is the first calculation of the static $B_{B}$ parameter that
includes dynamical fermions.
The method used is described in references~\cite{Christensen95a} and~\cite{Ukqcd95a}. The same set 
of smearing functions used in the $f_{B}^{static}$ calculations
was used to smear the quarks in the external mesons.
%
%
%
%
%
\begin{figure}
\vspace{-0.3in}
\vbox{\hskip -0.1in\epsfxsize=3in \epsfbox[0 0 4096 4096]{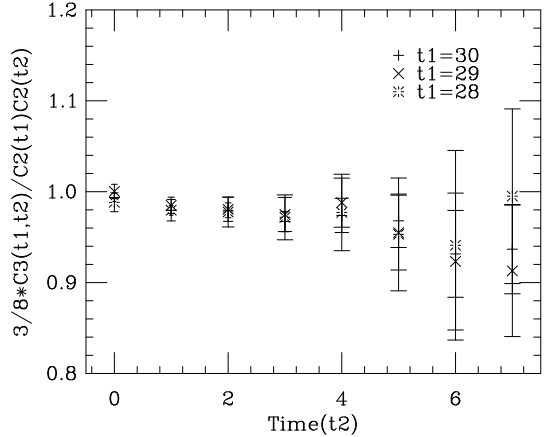} }
\vskip -0.5in
  \caption{$B_{L}$ parameter as a function of time slice, for
$\kappa = 0.156$, and source $\exp{(-0.4 * \rrmod)}$}
\label{fig:bfig_ps}
\vskip -0.35in
\end{figure}
Fig.~\ref{fig:bfig_ps} shows the static $B_{L}$ operator 
as a function of time.
In Table~\ref{tb:Bparam} we show some preliminary results for static $B_{B}$
parameter. The $\chi^{2}/dof$ for all the fits in
Table~\ref{tb:Bparam} were all close to 0.5. The errors are
statistical only; the systematic errors due to the choice of
fit range
are larger than the statistical errors. The errors should be reduced
when we include the additional smearing functions in our analysis.

We used the Kentucky group's~\cite{Christensen95a} organization of perturbation theory to
find the required linear combination of operators to calculate 
$B_{B}(m_{B})$. However we omitted some
next to leading order $\log \mu /m $ terms.
The missing terms have only recently been 
calculated~\cite{Buchalla96a}
and are a small effect.
%
%
\begin{table}[thb]
\caption{Static $B_{B}$ parameter results for $\MILCbeta = 5.6 $, $am_{sea} = 0.01$, using the
source $\exp{(-0.4 * \rrmod)}$, fixing timeslice 29, and fitting times 2 to 6.
}
\label{tb:Bparam}
  \centering
\begin{tabular}{cccc} \hline
  & $0.156$ & $0.158 $  & $0.159$  \\ \hline
$B_{L}$ & $0.98(1)$ & $0.99(2)$ & 1.00(2)  \\
$B_{R}$ & $0.96(1)$ & $0.94(2)$ & $0.94(2)$  \\
$B_{N}$ & $1.01(2)$ & $1.01(2)$ & $1.01(2)$  \\
$-\frac{8  }{5 } B_{S}$ & $1.00(1)$ & $1.00(1)$ & $1.01(2)$ \\ \hline
$ B_{B}(m_{B})$ & 0.91(2) & 0.92(2) & 0.93(2) \\ \hline
\end{tabular}
\vspace{-0.2in}
\end{table}
Chirally extrapolating the $B_{B}(m_{B})$ results to 
$\kappa_{c} = 0.16103$ and converting the results to the
one loop RG invariant $\hat{B}_{B}$ parameter, we obtain the
preliminary value of $\hat{B}_{B}= 1.31^{+3}_{-4}$ (statistical error only).
This result is consistent with
the values obtained from quenched simulations using Wilson 
fermions~\cite{FLYNN96a}~\cite{Christensen95a}.
\section{LATTICE HQET}
To calculate the form factors of the semi-leptonic decays
of the $B$ meson, we want to follow a strategy similar 
to the one used in the $f_{B}$ simulations, except that
we will combine the results of heavy quark effective field
theory (HQET) with the analogous propagating quark calculations,
to allow the final results to be interpolated to the $B$ meson mass.

As a ``warm up exercise'' we studied the two point function of a
HQET-light meson. This exercise allows us to investigate the smearing of the
quarks in the $B$ meson and to study the nonperturbative renormalization
of the velocity (a peculiarity of lattice HQET) -- both of which are
important 
prerequisites to the calculation of form factors.

We have implemented the  HQET propagator equation
introduced by Mandula and Ogilvie \cite{Mandula92a}.
The two-point function for an HQET-light meson
at finite residual momentum $p$ and bare velocity $v$ is 
\begin{eqnarray}
C(\p,t;v) & = & \sum_{\x} \sum_{\rr} f(\rr) e^{ i p .x } \nonumber \\
& & \langle 
   \overline{b_{v}}( \underline{r},0 ) \gamma_{5}  q( \underline{0},0 )
   \overline{q}( \x , t ) \gamma_{5}  b_{v}( \x ,t)
\rangle 
\nonumber
\end{eqnarray}
When there is no excited state contamination, the correlator has the form
\begin{equation}
C(\p,t;v) = Z_{f}^{v} Z_{L}^{v} e^{ - E(p,v^{R}) t }
\end{equation}
where $Z_{f}^{v} $ and $Z_{L}^{v} $ are the smeared and local matrix
elements. The dispersion relation for an HQET-light meson is \cite{Hashimoto95a}
\begin{equation}
E(\p,v^{R}) = \frac{ E_{sim} } { v_{0}^{R} }  + \frac{ \underline{v^{R}}  . \hat{ \p } } {
v_{0}^{R}  }
\label{eq:HQETdisp}
\end{equation}
where $E_{sim}$
is equal to the energy obtained in the static simulation. 
Here $v^{R}$ is the renormalized velocity.

\begin{figure}
\vbox{\hskip -0.2in\epsfxsize=3.0in \epsfbox[0 0 4096 4096]{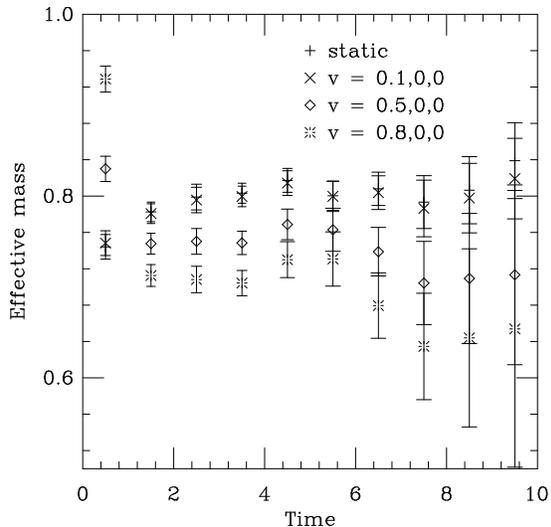} }
\vskip -0.5in
\caption{Effective mass plot for the HQET-light two point function,
with source $\exp{(-0.67 * \rrmod)}$}
\label{fig:hqet_ps}
\vskip -0.3in
\end{figure}

As a pilot study we generated 20 correlators at $\MILCbeta = 5.445$ with a
light quark
kappa value of 0.160. So far, a single exponential source, as
in the static simulations, was used. Fig.~\ref{fig:hqet_ps} shows the
effective mass plots for a HQET-light meson at zero residual
momentum with bare velocities only in the $x$ direction of
0.1, 0.5, and 0.8; the static effective mass plot is also included
(equivalent to zero velocity). The single exponential produces almost
usable plateaus for all three velocities -- the plateaus 
should improve when we use variational smearing. The signal to noise
ratio
does not decrease rapidly with increasing velocity.
%
%
%
 The renormalization of the velocity is caused by the breaking of the 
Lorentz symmetry by the lattice. This renormalization is required in
the calculation of form factors such as the Isgur-Wise function~\cite{Aglietti94a}.
Our preferred way to extract the renormalization is to use the
dispersion
relation in Eq~(\ref{eq:HQETdisp}) with the energy taken from
simulations.
This approach seems to us~\cite{Hashimoto95a}~\cite{Draper96a0} to be simpler than the one used by Mandula and 
Ogilvie~\cite{Mandula96a}.

An estimate of the velocity renormalization can be obtained from the
data in
Fig.~\ref{fig:hqet_ps}, using naive fits to the effective masses
and using Eq~(\ref{eq:HQETdisp})  with zero residual momentum. For
the
bare velocity of 0.5 (0.8) in the $x$ direction, the renormalized velocity was
approximately eighty (sixty five) percent of the bare velocity.
With higher statistics and better
smearing a more sophisticated analysis will be done. The perturbative
calculation of
Mandula and Ogilvie~\cite{Mandula96a} also gave a renormalized velocity smaller than
the bare velocity.

%
%

This work is supported in
part by the U.S. Department of Energy and the NSF
We would like to thank Terry Draper for discussions on HQET and $B_{B}$,
and Joe Christensen for providing us with the coefficients for the 
$B_{B}$ parameter calculation.
The
runs are being done on the Cornell Theory Center's
SP2, and on the 512 node Paragon at the Oak Ridge Center for
Computational Science.


\begin{thebibliography}{10}

\bibitem{FLYNN96a}
J. Flynn, these proceedings.

\bibitem{MILC}
{(MILC collaboration) C. Bernard} {\it et~al.}, {\sl {\sl Nucl.\ Phys.\ B}
  ({\sl Proc.\ Suppl.\ })} {\bf 42},  388  (1995);
{(MILC collaboration) C. Bernard} {\it et~al.}, {\sl {\sl Nucl.\ Phys.\ B}
  ({\sl Proc.\ Suppl.\ })} {\bf 47},  459  (1996);
MILC collaboration (presented~by C.~Bernard), these proceedings.

\bibitem{Draper94a0}
T. Draper and C. McNeile, {\sl {\sl Nucl.\ Phys.\ B} ({\sl Proc.\ Suppl.\ })}
  {\bf 34},  453  (1994).

\bibitem{Bitar92a}
K. Bitar {\it et~al.}, {\sl Phys. Rev. D} {\bf 46},  2169  (1992).

\bibitem{AliKhan96a}
A.~Ali Khan {\it et~al.}, preprint hep-lat/9607004, 1996.

\bibitem{Christensen95a}
{J. Christensen, T. Draper} and {C. McNeile}, to be published; and these
  proceedings.

\bibitem{Ukqcd95a}
{(UKQCD Collaboration) A.K. Ewing} {\it et~al.}, hep-lat/9508030, 1995.

\bibitem{Buchalla96a}
G. Buchalla, hep-ph/9608232, 1996;
{M. Ciuchini and E. Franco} and V. Gim\'enez, hep-ph/9608204, 1996.

\bibitem{Mandula92a}
J. Mandula and M. Ogilvie, {\sl Phys. Rev. D} {\bf 45},  2183  (1992).

\bibitem{Hashimoto95a}
S. Hashimoto and H. Matsufuru, preprint hep-lat/9511027, 1995.

\bibitem{Aglietti94a}
U. Aglietti, {\sl Nucl. Phys.} {\bf B421},  191  (1994).

\bibitem{Draper96a0}
T. Draper and C. McNeile, {\sl {\sl Nucl.\ Phys.\ B} ({\sl Proc.\ Suppl.\ })}
  {\bf 47},  429  (1996); T. Draper and C. McNeile in preparation.

\bibitem{Mandula96a}
J. Mandula and M. Ogilvie, preprint hep-lat/9602004, 1996;
and these proceedings.

\end{thebibliography}

\end{document}